\documentclass[aps,prl,twocolumn,superscriptaddress,showpacs]{revtex4}
\usepackage{graphicx}
\usepackage{dcolumn}
\usepackage{bm}

\begin{document}

\title{Effect of antiferromagnetic spin correlations on lattice distortion and charge ordering in  Pr$_{0.5}$Ca$_{1.5}$MnO$_{4}$}

\author{Songxue~Chi}
\affiliation{Department of Physics and Astronomy, The University of Tennessee,
Knoxville, Tennessee 37996-1200, USA}

\author{F.~Ye}
\affiliation{Oak Ridge National Laboratory, Oak Ridge, Tennessee 37831-6369, USA}

\author{Pengcheng~Dai}
\email{daip@ornl.gov}
\affiliation{Department of Physics and Astronomy, The University of Tennessee,
Knoxville, Tennessee 37996-1200, USA}
\affiliation{Oak Ridge National Laboratory, Oak Ridge, Tennessee 37831-6369, USA}

\author{J.~A.~Fernandez-Baca}
\affiliation{Department of Physics and Astronomy, The University of Tennessee,
Knoxville, Tennessee 37996-1200, USA}
\affiliation{Oak Ridge National Laboratory, Oak Ridge, Tennessee 37831-6369, USA}

\author{Q.~Huang}
\affiliation{NIST Center for Neutron Research, National Institute of Standards
and Technology, Gaithersburg, Maryland 20899-6102, USA}

\author{J.~W.~Lynn}
\affiliation{NIST Center for Neutron Research, National Institute of Standards
and Technology, Gaithersburg, Maryland 20899-6102, USA}

\author{E.~W.~Plummer}
\email{eplummer@ornl.gov}
\affiliation{Department of Physics and Astronomy, The University of Tennessee,
Knoxville, Tennessee 37996-1200, USA}
\affiliation{Oak Ridge National Laboratory, Oak Ridge, Tennessee 37831-6369, USA}

\author{R.~Mathieu}
\affiliation{Spin Structure Project (ERATO), Japan Science and Technology Corporation, Advanced Industrial Science and Technology Center 4, Tsukuba 305-8562, Japan}

\author{Y. Kaneko}
\affiliation{Spin Structure Project (ERATO), Japan Science and Technology Corporation, Advanced Industrial Science and Technology Center 4, Tsukuba 305-8562, Japan}

\author{Y.~Tokura}
\affiliation{Spin Structure Project (ERATO), Japan Science and Technology Corporation, Advanced Industrial Science and Technology Center 4, Tsukuba 305-8562, Japan}
\affiliation{Department of Applied Physics, University of Tokyo 113-8656, Japan}

\date{\today}

\begin{abstract}
We use neutron scattering to study the lattice and magnetic structure of the layered half-doped manganite Pr$_{0.5}$Ca$_{1.5}$MnO$_4$.  On cooling from high temperature, the system first becomes charge- and orbital- ordered (CO/OO) near $T_{CO}=300$~K and then develops checkerboard-like antiferromagnetic (AF) order below $T_{N}=130$ K. At temperatures above $T_{N}$ but below $T_{CO}$ ($T_N<T<T_{CO}$), the appearance of short-range AF spin correlations suppresses the CO/OO induced orthorhombic strain, contrasting with other half-doped manganites, where AF order has no observable effect on the lattice distortion.  These results suggest that a strong spin-lattice coupling and the competition between AF exchange and CO/OO ordering ultimately determines the low-temperature properties of the system.\end{abstract}


\maketitle


Understanding the competition and coupling between charge, lattice, and 
spin degrees of freedom in doped transition metal oxides continues to be one of the
most profound intellectual challenges in modern condensed matter physics since  
the discovery of high-transition temperature (high-$T_c$) superconductors and the 
colossal magnetoresistance (CMR)  
manganese oxides \cite{review}. The complexity of transition metal oxides is directly responsible
for their tunability and the balance between different competing phases
can produce large changes in the physical properties.  For example,  
superconductivity in the high-$T_c$ superconductors
La$_{2-x}$Ba$_{x}$CuO$_4$ becomes
drastically suppressed at the doping level of 
$x=1/8$ due to the spin and charge phase separation 
(the so-called ``striped'' phase), where charge ordering (CO) establishes 
a template at a higher temperature to be followed by antiferromagnetic (AF) 
stripe order at a lower temperature \cite{tranquada,fujita}. The low-temperature AF phase
has little or no influence on the already established CO phase because of its low energy scales.
Similarly, the long-range AF order in the parent compounds of high-$T_c$ copper oxides 
is charaterized by spin-only antiferromagnetism and has little or no effect on the underlying
lattice \cite{review}. These results suggest that spin-lattice coupling is weak in 
high-$T_c$ copper oxides.

In the case of CMR manganites $A_{1-x}A^{\prime}_{x}$MnO$_3$ 
(where $A$ and $A^\prime$ are trivalent rare-
and divalent alkaline-earth ions respectively), the competition between charge, lattice, and spin degrees of freedom
can be delicately balanced to form a variety of ground states \cite{review}.
Before doping any holes into the system, the parent compound such as LaMnO$_3$
has an insulating ground state, where the Mn$^{3+}$ spins 
order in the A-type AF structure \cite{Wollan,Goodenough}. 
For hole-doping level around $x=0.3$ by substituting trivalent 
La$^{3+}$ with divalent Ca$^{2+}$, La$_{1-x}$Ca$_x$MnO$_3$ becomes a metallic ferromagnet  
with a CMR effect near the Curie temperature $T_C$. The formation of long-range ferromagnetic
order at $T_C$ also induces a large lattice distortion, suggesting a strong spin-lattice coupling \cite{dai96}.
Upon increasing the doping level to $x=0.5$, La$_{0.5}$Ca$_{0.5}$MnO$_3$
changes again into an AF insulating phase but with a CE-type AF structure \cite{Wollan}.
Here, equal amounts of Mn$^{3+}$ and Mn$^{4+}$
distribute alternately in the MnO$_2$ plane of
La$_{0.5}$Ca$_{0.5}$MnO$_3$, forming a checkerboard CE-type 
pattern as schematically depicted in Figure 1(a) \cite{Wollan,Goodenough}.
Although the CE-type AF order disappears on warming above the N{\'{e}}el temperature 
$T_N$, the system is still charge and orbitally ordered (CO/OO).  Such CO/OO order     
is strongly coupled to the lattice and induces an orthorhombic distortion that only disappears
at temperatures well above CO/OO ordering temperature $T_{CO}$.

For example, in the three-dimensional nearly half-doped perovskites
La$_{0.5}$Ca$_{0.5}$MnO$_3$ \cite{Radaelli}, Pr$_{0.5}$Ca$_{0.5}$MnO$_3$ \cite{jirak,aladine},
and Pr$_{0.55}$(Ca$_{0.8}$Sr$_{0.2}$)$_{0.45}$MnO$_3$ \cite{ye}, the CO/OO ordered lattice first established
slightly below room temperature is followed by a CE-type AF order around 130~K [Fig.~1(a)].
X-ray and neutron diffraction experiments have
shown that the materials exhibit a tetragonal to orthorhombic phase transition near $T_{CO}$. 
Furthermore, the orthorhombicity increases with decreasing
temperature and shows no anomalies across the CE-type AF phase
transition \cite{Radaelli,jirak,aladine,ye}.  These results thus suggest that CO/OO order is strongly
coupled to the lattice and  there is a weak spin-lattice coupling.  As a consequence, CO/OO ordering in
half-doped perovskites 
may have a larger energy scale than the low temperature magnetic order.
For the single layer half-doped
manganites such as La$_{0.5}$Sr$_{1.5}$MnO$_4$ (LSMO), a similar behavior 
is also observed. Here, the material
exhibits a tetragonal to orthorhombic phase transition at the CO/OO
temperature of 230~K and then orders antiferromagnetically with a
CE-structure below about 120~K \cite{sternlieb,Larochelle2}. The lattice distortion and
orthorhombicity of LSMO show no anomalies below the AF phase transition. 
Therefore, it appears that 
CO/OO order in doped transition metal oxides generally is strongly coupled to the lattice,
while the low-temperature 
magentic order has no influence on CO/OO ordering.

Although CO/OO order in doped manganites may have a stronger coupling to the lattice than that of 
the AF order, its microscopic origin is still unclear. Theoretically, CO/OO order established at higher
temperatures may actually have a purely magnetic spin origin \cite{solovyev}; arise from a competition between
the kinetic energy of the electrons and the magnetic exchange energy
\cite{brink}, due to a tendency of the Jahn-Teller distorted Mn$^{3+}$ ions
to maximize their relative distances to gain electron kinetic
energy \cite{yunoki}, or come from a purely Coulomb interaction without
invoking magnetic interactions \cite{mutou,khomskii}.  
In general, charge ordering in half-doped manganites is intimately related to the
orbital ordering, where the orbitals of $e_g$ electrons on Mn$^{3+}$ sites form zigzag
ferromagnetic chains that order antiferromagnetically [Fig. 1(a)] \cite{murakami,Dhesi}.
One way to
sort out the relationship between CO/OO and CE-type AF order is to
carry out systematic measurements on $A_{0.5}A^{\prime}_{0.5}$MnO$_3$ or 
layered $A_{0.5}A^{\prime}_{1.5}$MnO$_4$ with
different $A$ and $A^\prime$ ionic sizes. Decreasing the ionic size
at $A$ and $A^\prime$ sites in half-doped manganites 
increases the buckling of the MnO$_6$ octahedra and therefore the
lattice distortion of the perovskite.
For three-dimensional
$A_{0.5}A^{\prime}_{0.5}$MnO$_3$, replacing Sr in
Pr$_{0.5}$Sr$_{0.5}$MnO$_3$ ($T_{CO}=150$~K) by the smaller Ca to
form Pr$_{0.5}$Ca$_{0.5}$MnO$_3$ ($T_{CO}=260$~K) moderately
enhances the CO/OO ordering temperature, but dramatically increases the
magnitude of the magnetic field (from 5 T for
Pr$_{0.5}$Sr$_{0.5}$MnO$_3$ to 27 T for Pr$_{0.5}$Ca$_{0.5}$MnO$_3$)
needed to suppress CO/OO \cite{tokura}.
These results suggest that
CO/OO ordering is more stable for manganites with smaller ionic size and larger
lattice distortion; and has an energy scale larger than that of the magnetic
exchange.  Since single crystals of 
three-dimensional $A_{0.5}A^{\prime}_{0.5}$MnO$_3$
with the CE-type AF structure are unavailable, we study 
$A_{0.5}A^{\prime}_{1.5}$MnO$_4$ with different $A$ and $A^\prime$ ionic sizes.

\section {Results}
Here we present neutron scattering results on
Pr$_{0.5}$Ca$_{1.5}$MnO$_4$ (PCMO), a single layer manganite with smaller 
average $A$ and $A^\prime$ site ionic radius and 
larger lattice distortion than that of LSMO \cite{distortion}.  We chose to study PCMO
in order to determine the effect of the lattice distortion on the CO/OO and AF
phase transitions.  Since CO/OO is not affected by
CE-type AF order in LSMO \cite{sternlieb,Larochelle2}, 
one would expect that CO/OO becomes more robust when the larger (La,Sr) ions in LSMO are
replaced by smaller (Pr,Ca) in 
PCMO.  Surprisingly, we find that the development of short-range
AF spin correlations in the MnO$_2$ plane of PCMO significantly 
affects the CO/OO-induced lattice distortion and
 reduces the orthorhombicity of the system below $T_N$.
Our results thus indicate the presence of a strong spin-lattice
interaction, suggesting that antiferromagnetism can 
reduce the CO/OO-induced orthorhombic strain and thus 
compete with the CO/OO ordering. 

\begin{figure}
\includegraphics[width=3.2in]{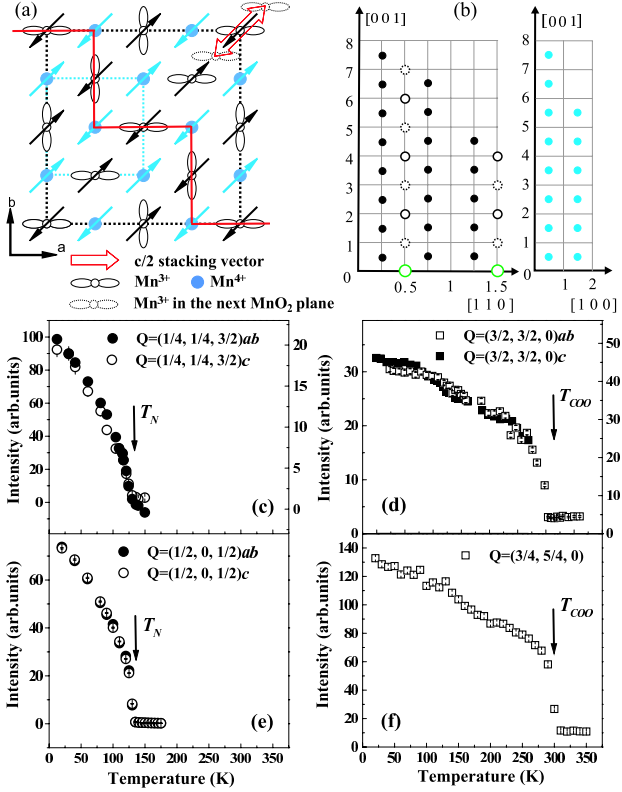}
\caption{\label{fig1} Structural scatterings and their temperature dependence.
(a) Schematic view of the CE-type AF 
ordering in the $\rm MnO_2$-plane. The black dashed line represents the
periodicity of the unit cell for the Mn$^{3+}$ sublattice, and the
blue dashed line shows that of the Mn$^{4+}$ sublattice. 
Possible spin arrangements in the $c/2$ stacking layers are marked
by red arrows. The directions
of Mn$^{3+}$ orbitals form zigzag ferromagnetic chains (red line) that order
antiferromagnetically. (b) The
observed nuclear peaks (black open circles), CO-OO-induced
superlattice peaks (green open circles) and magnetic ordering (solid 
circles) in reciprocal space. The dotted open circles represent the observed weak nuclear peaks that are 
disallowed by orthorhombic symmetry, indicating that the symmetry is lower than orthorhombic. Temperature dependence of the AF peak
intensity from (c) (1/4,1/4,3/2); (e) (1/2,0,1/2) and temperature
dependence of CO-OO peak intensity from (d) (3/2,3/2,0) and from (f)
(3/4,5/4,0).}
\end{figure}

\begin{figure}
\includegraphics[width=3.2in]{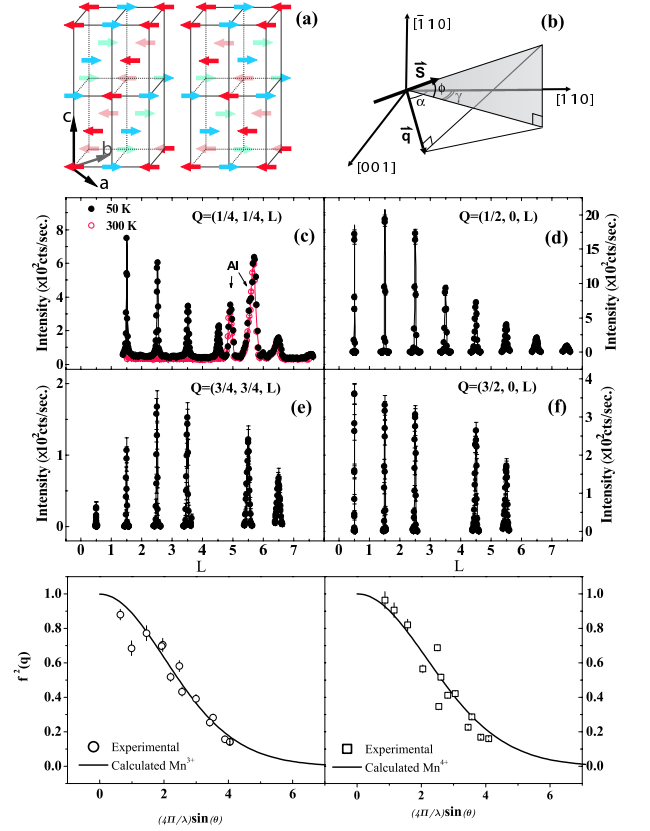}
\caption{\label{fig2} The magnetic structure determination of PCMO. (a) 
Two possible spin arrangements for the Mn$^{3+}$ sublattice as obtained
from Rietveld analysis of the HRNPD data and fits to single crystal
integrated intensities at different positions. (b) The geometrical relationship
between the Mn$^{3+}$ spin and the MnO$_2$ plane.
(c) Scattering data along ${\bf q}$=(1/4,1/4,L) at $T=50$ and 300~K, respectively. 
Panels (d), (e) and
(f) show the $\theta-2\theta$ scans for ${\bf q}$=(3/4,3/4,L),
(1/2,0,L) and (3/2,0,L) that are projected to the [0,0,L] direction. The
intensities of observed magnetic peaks are fit to the generic
magnetic form factor for (g) Mn$^{3+}$ and (h) Mn$^{4+}$
ions.}
\end{figure}

We grew single crystals of PCMO using the traveling solvent floating zone technique.
At room temperature, PCMO has the orthorhombic
structure with lattice parameters $a_o=5.380$ \AA\, $b_o=5.404$ \AA\ and
$c_o=11.831$ \AA\ (space group {\it bmab}). For simplicity, we use
the tetragonal unit cell for the triple-axis measurements and label the
momentum transfers ${\bf q}=(q_x, q_y, q_z)$ 
as $(h, k, l)=(q_xa/2\pi, q_ya/2\pi, q_zc/2\pi)$ in
reciprocal lattice units (rlu), where $a=(a_o+b_o)/2\sqrt{2}=3.814$ \AA.

Because one expects PCMO to behave similarly to LSMO, 
we first probe the low temperature magnetic and superlattice peaks associated with
the CE-type AF structure and CO/OO state. Figs 1(d) and (f) show
the temperature dependence of the ${\bf q}=(3/2,3/2,0)$
and ${\bf q}=(3/4,5/4,0)$ structural superlattice peaks,
respectively.  Below $\sim$310~K, a structural phase transition
associated with the CO/OO ordering occurs, consistent with the large increase in
resistivity from transport measurements \cite{Ibarra}.  Figs
1(c) and (e) show the temperature dependence of the AF Bragg peaks at 
${\bf q}=(1/4,1/4,3/2)$ and
${\bf q}=(1/2,0,1/2)$, corresponding to the Mn$^{3+}$ and
Mn$^{4+}$ of the CE-type AF structure in Fig.~1(a),
respectively. The system develops AF order below 130~K,  
consistent with the results of 
bulk transport measurements \cite{mathieu} and similar to
other half-doped manganites
\cite{Wollan,jirak,Radaelli,aladine,ye,sternlieb,Larochelle2}.

\begin{figure}
\includegraphics[width=3.2in]{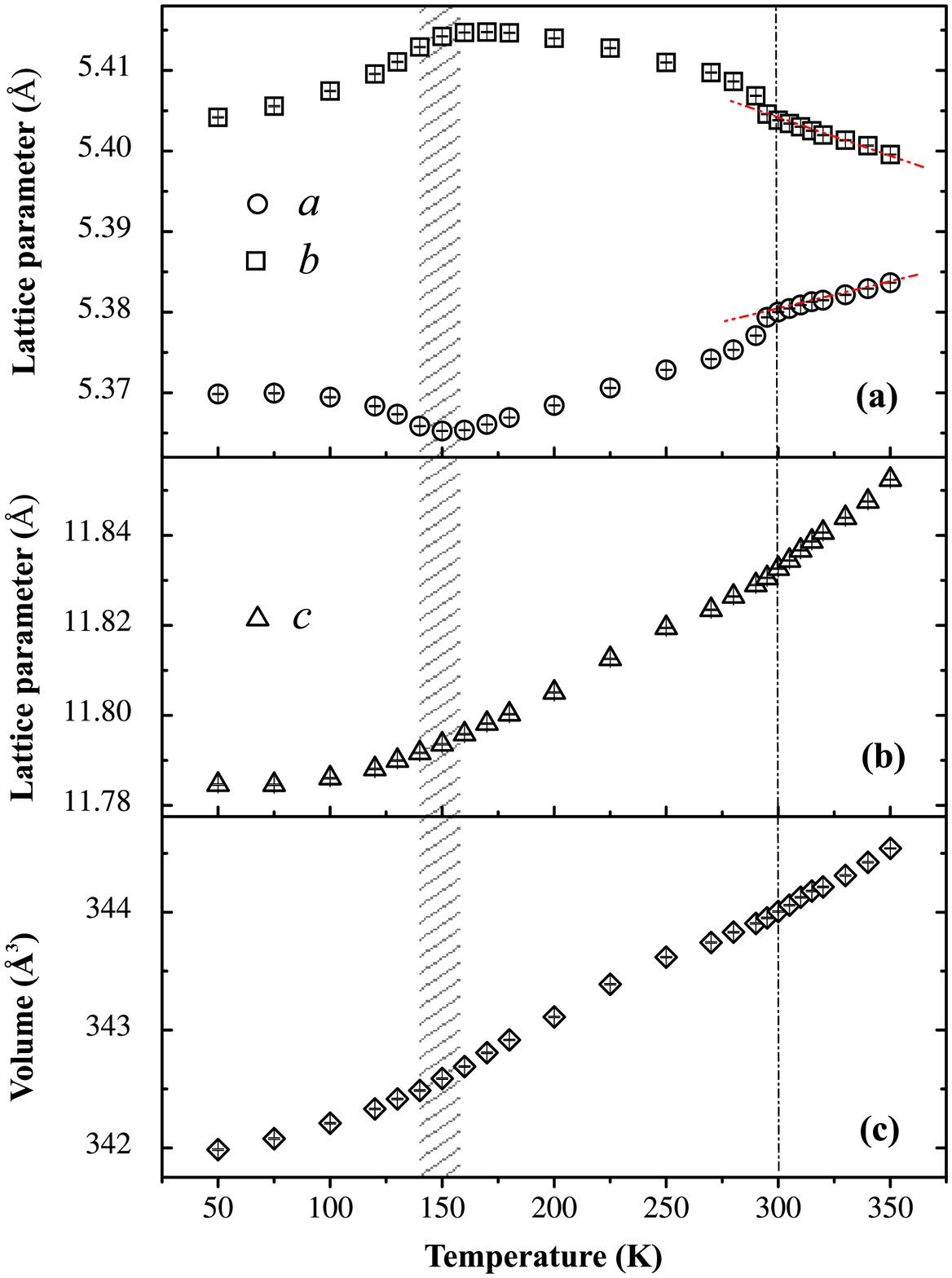}
\caption{\label{fig3} Temperature dependence of lattice parameters
and unit cell volume. The dashed line near 300~K marks the CO-OO
transition temperature $T_{CO}$. While
the in-plane $a$ and $b$ lattice parameters show clear anomalies
around $T_{CO}$ and $T_N$, the $c$-axis lattice parameter changes 
smoothly across both
transitions. 
The dash-dotted lines in panel
(a) are guides to the eye.}
\end{figure}

To determine the low-temperature magnetic structure of PCMO, we made
extensive surveys of reciprocal space and found that the allowed
magnetic peaks are characterized by wavevectors ${\bf q}=(2n+1/4, 2n+1/4,
l)$ and $(2n+1/2, 0, l)$ with $n$ and $l$ being integers and
half-integers, respectively [Fig.~1(b)].  Figure 2
summarizes scans along the $l$ direction for the $(2n+1/4, 2n+1/4,
l)$ (corresponding to the Mn$^{3+}$ sites) and $(2n+1/2, 0, l)$ (the
Mn$^{4+}$ sites) Bragg positions.  The $l=m+1/2$ ($m=0,1,2,\cdots$)
peaks are clearly magnetic because they disappear above the N{\rm
\'{e}}el temperature [Fig.~2(c)]. 
We note that the $c$-axis correlation length in PCMO is resolution-limited and long-ranged,
in contrast to the short-range $c$-axis correlations in LSMO \cite{sternlieb}.
 Magnetic structure factor
calculations indicate two possible spin stackings of successive
MnO$_2$ layers along the $c$-axis direction. As depicted in
Fig.~1(a), spins in the $c/2$ MnO$_2$ layer simply shift from those in
the $c=0$ layer by $(a/2,a/2,c/2)$ or $(-a/2,-a/2,c/2)$. The stacking arrangements of Mn$^{3+}$ sublattice are also shown in Fig.~2(a). The
resulting magnetic structure allows both $(2n+1/4, 2n+1/4, l)$ and
$(2n+1/2, 0, l)$ peaks. There is no evidence of magnetic
peaks at $l$-even $(2n+1/4,2n+1/4,l)$ positions [Fig.~2(c)] as observed 
in LSMO
\cite{sternlieb}.  The temperature dependence of the order
parameters for the $(1/4,1/4,3/2)$ [Fig.~1(c)] and $(1/2,0,1/2)$ 
[Fig.~1(e)] peaks show that the Mn$^{3+}$ and Mn$^{4+}$ networks enter the
AF long-range ordered states simultaneously at $T_N \sim$130~K.

We measured the radial and transverse scans of all
observed magnetic peaks. The product of the longitudinal peak width
in full-width-half-maximum (FWHM) and the integrated intensity of
the rocking curve was used as the total intensity of a Bragg peak. The
observed intensity of a magnetic Bragg peak should be proportional
to  
$
I\propto{|{F_M}(\bf{q})|^2}/{\sin(2\theta)}
$, where $\theta$ is the
scattering angle.  The magnetic structure factor ${F_M}$ is \begin{equation}
F_M({\bf{q}}) = \sum_{j}f({\bf{q}})_j {\bf{q}}
\times({\bf{M}}_j\times {\bf{q}} ) e^{i {\bf{q}} \cdot {\bf{r}}
} e^{-W_j},
\end{equation} where $f(\bf{q})_j$, $\bf{M}_j$ and $e^{-W_j}$ are the magnetic
form factor, the spin  moment of the $j$-th ion and Debye-Waller factor 
respectively. 

\begin{figure}
\includegraphics[width=3.2in]{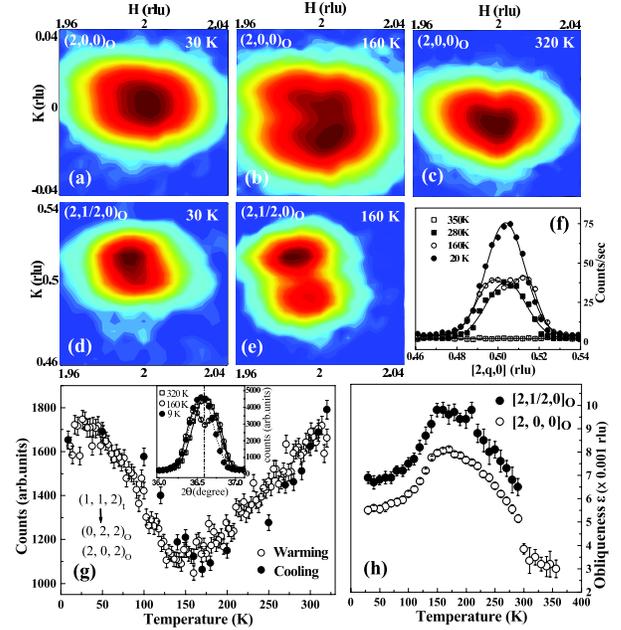}
\caption{\label{fig3} Strong spin-lattice coupling near the magnetic transition temperature. (a-c) Mesh-scans around the nuclear Bragg peak (2,0,0)$_O$ [in orthorhombic notation] at $T=30$, 160 and
320~K.  (d and e) The corresponding mesh-scans around
CO-OO induced superlattice peak (2,1/2,0)$_O$ at 30 and 160~K. (f) wavevector 
scans of the same CO-OO peak at selected temperatures.
(g) Temperature dependence of the peak intensity from powder monitored at
2$\theta$=36.61$^{\circ}$, which corresponds to (1,1,2)$_t$
structural peak in tetragonal notation. The inset shows the
splitting of the (1,1,2)$_t$ peak [the actual (0,2,2)$_O$ and
(2,0,2)$_O$ in orthorhombic symmetry] becomes much more prominent at
160~K and recovers back to one peak at low temperature. (h)
Temperature dependence of the obliqueness, the separation between the center of
the split peaks in reciprocal space, for (2,0,0)$_O$
and (2,1/2,0)$_O$.}
\end{figure}

\begin{figure}
\includegraphics[width=3.2in]{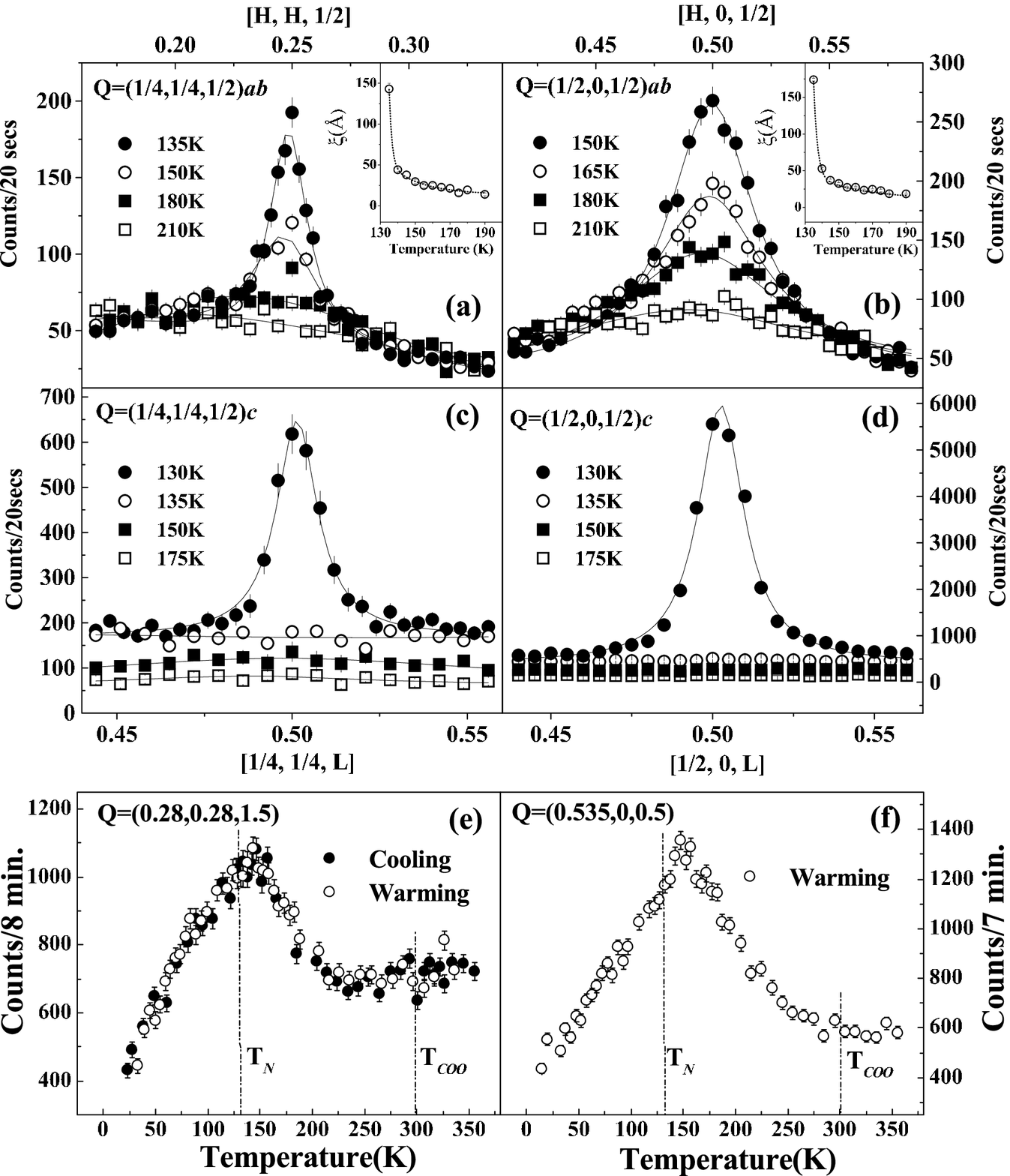}
\caption{\label{fig5} Crossover from two-dimensional AF fluctuations to 
three-dimensional AF order. Wavevector scans of AF scattering from the
$\rm Mn^{3+}$ sublattice near ${\bf q}=(1/4,1/4,1/2)$ (a) within 
Mn-O plane and (c) along the inter-plane direction. Similar scans from the
$\rm Mn^{4+}$ sublattice near (1/2,0,1/2) are presented in panels (b)
and (d). Insets shows the evolution of magnetic correlation lengths
above the long range AF order temperature $T_N=130$~K.
(e) Temperature profiles of short-range magnetic scattering measured
at (e) ${\bf q}=(0.28,0.28,3/2)$ and at (f) ${\bf q}=(0.535,0,1/2)$.
Those wavevectors have been chosen to avoid the
contamination from the magnetic Bragg peaks.}
\end{figure}

In the case of the Mn$^{3+}$ spin network, the integrated intensities of
$(2n+1/4, 2n+1/4, l)$ peaks depend on $\alpha$, $\phi$ and $\gamma$,
where $\alpha$ is the the angle between wave vector $\bf q$ and the
MnO$_2$ plane, $\phi$ is the angle between the moment direction and
the $[1,1,0]/[0,0,1]$ plane, and $\gamma$ is the angle between the
projection of the spins in the $[1,1,0]/[0,0,1]$ plane and the
MnO$_2$ plane, as depicted in Figure 2(b).  By fitting
the integrated intensities of $(2n+1/4, 2n+1/4, l)$ peaks  as a
function of $\alpha$, $\phi$ and $\gamma$, we find that the best fit
for the Mn$^{3+}$ form factor in Fig. 2(g) requires both $\phi$ and $\gamma$ to
be zero, indicating that the Mn$^{3+}$ spins are in the MnO$_2$
basal plane and along the $[1,1,0]$ direction [Figs.~1(a) and 
Fig. 2(a)]. Similarly, the moment direction for Mn$^{4+}$ spins along the
$[1,1,0]$ direction also gives the best fit [Fig.~2(h)].
Independent Rietveld analysis of the magnetic structural data on powder
samples confirms that the magnetic structure has dimensions of
$a_o \times 2b_o \times 2c_o$ (where $a_o=5.37$~\AA,
$b_o=5.40$~\AA, and $c_o=11.78$~\AA\ at low temperature) for the Mn$^{3+}$ magnetic
sublattice and $2a_o \times 2b_o \times 2c_o$ for Mn$^{4+}$
sublattice.  Although the proposed spin directions of PCMO is different
from that of LSMO, where spins are aligned along the $[1,2,0]$
direction in the MnO$_2$ basal plane, the presence of 
impurity and minority phase in LSMO makes the conclusive 
magnetic strucutral determination difficult \cite{sternlieb}.

As PCMO is cooled from 350~K, the orthorhombicity of its structure 
increases with decreasing temperature and shows a clear enhancement
of the orthorhombic strain 
 around the charge ordering temperature $T_{CO}$ to
accommodate the establishment of orbital ordering. Figure 3 shows
the temperature dependence of the lattice parameters and unit cell
volume obtained from Rietveld analysis of the neutron powder
diffraction data. While an enhancement of the orthorhombic structure
near $T_{CO}$ is expected, similar to that of other
half-doped manganites
\cite{Radaelli,jirak,aladine,ye,sternlieb,Larochelle2}, the
orthorhombicity of PCMO mysteriously becomes smaller below
$\sim$150~K, at temperature 20~K above the $T_N$ of the system
(Fig.~3).  To demonstrate this more clearly, we carried out detailed
studies of the $(1,1,2)$ Bragg peak at temperatures 30~K$<T_N$,
$T_N<$160~K$<T_{CO}$, and $T_{CO}<$320 K [Fig. 4(g)]. Below $T_{CO}$, the
$(1,1,2)$ peak at 2$\theta$=36.61$^{\circ}$ starts to broaden with
reduced peak intensity, and then splits into two peaks [indexed as 
$(0,2,2)_O$ and $(2,0,2)_O$ in orthorhombic notation] at
$T\sim$150~K as shown in the inset of Fig.~4(g).  As the
temperature continues to drop, the split peaks merge back into one
at low temperature. The temperature dependence of the $(1,1,2)$ peak
intensity shows a continuous drop for  
$T<300$ K and then the recovery below $T\sim$150~K [Fig.~4(g)].  

Figs. 4(a-f) summarize mesh scans in reciprocal space near the
fundamental Bragg $(2,0,0)_O$ and charge ordering $(2,1/2,0)_O$
positions in the orthorhombic symmetry at low, intermediate, and high temperatures obtained
on single cyrstals of PCMO. The $(2,0,0)_O$
peak first broadens and then splits along the transverse direction at 160~K. On
further cooling to 30~K, the split peaks become one again [Fig.
4(a)]. Figs. 4(d-f) show that the $(2,1/2,0)_O$ CO-OO peak, 
which is equivalent to the $(3/4,5/4,0)$ peak in tetragonal notation, exhibits similar
behavior: broadens and splits between $T_N$ and $T_{CO}$, and
emerges back to one below $T_N$. To quantitatively determine the
degree of orthorhombicity, we plot in Fig.~4(h) the temperature
dependence of the separation between the centers of split peaks
$\epsilon$ in reciprocal space. Below $T_{CO}$ of 310~K, the
distortion increases dramatically. It continues to increase and
reaches its maximum around 150~K. Upon further cooling below $\sim$150
K (a temperature 20 K above $T_N$), the lattice distortion is
continuously suppressed, but still remains at the lowest probed
temperature of 20~K.  This anomalous lattice response near $T_N$ has
not been observed in LSMO or other half-doped manganite systems.
In these materials, the CO/OO induced lattice distortions  
do not exhibit noticeable anomaly across $T_N$ at 
lower temperatures \cite{Radaelli,jirak,Larochelle2}. We also note that the
suppression of orthorhombicity below $\sim$150~K in PCMO is not associated
with the melting of charge ordering as the integrated intensity of
CO peaks shown in Figs.~1(d) and 1(f) display no anomalies across
$T_N$.  This is different from bilayer perovskite manganites
\cite{Kimura,Argyriou}. 

The temperature dependence of AF peaks such as $(1/4,1/4,3/2)$
and $(1/2,0,1/2)$ shows a $T_N$ of 130~K for PCMO. Wavevector scans
within the MnO$_2$ plane and along the $c$-axis [Figs.~5(a)-5(d)]
show quite anisotropic correlations above $T_N$. 
Scans along the
$[h,h,1/2]$ and $[h,0,1/2]$ directions in the MnO$_2$ plane display 
the clear presence of two-dimensional short-range spin correlations
above $T_N$. Figure 5(a) suggests that the in-plane Mn$^{3+}$-Mn$^{3+}$
spin correlations are established at temperatures as high as 210~K,
while the inter-plane Mn$^{3+}$-Mn$^{3+}$ spin correlations are
turned on only below $T_N$ [Fig.~5(c)]. The spin correlations
between Mn$^{4+}$ ions behave similarly as shown in Figs.~5(b) and
4(d). The short-range AF spin correlations have been fit to a 
Lorentzian line shape as shown in the solid curves in Figs.~5(a) and
5(b).  Their linewidths decrease with decreasing temperature.  Below
$T_N$, the Lorentzian line shape is gradually taken over by a Gaussian
component indicating the development of long-range AF order. The
insets of Figs.~5(a) and 5(b) show the temperature dependence of the
in-plane spin-spin correlation lengths. While the correlation
lengths clearly diverge near $T_N$ as expected with the
establishment of the long-range AF order, there is no anomaly 
around $\sim$150~K. 
 
One way to determine the temperature dependence of the staggered
magnetic susceptibility is to track the scattering intensity at a
wavevector position slightly away from the magnetic Bragg peak (to
avoid the Gaussian component) but close enough to probe short-range
spin-spin correlations.  In a standard second order AF phase
transition, one would expect the staggered susceptibility to increase
with decreasing temperature, peak at the transition temperature and
then decrease below $T_N$.  Figs.~5(e) and 5(f) show the
temperature dependence of the scattering intensity at 
$(0.28, 0.28, 3/2)$ and $(0.535, 0, 1/2)$, which probe the
Mn$^{3+}$ and Mn$^{4+}$ spin-spin correlations, respectively. The
susceptibilities corresponding to Mn$^{3+}$ and Mn$^{4+}$ spin
correlations start to increase around 240~K.  They reach their
maxima at $\sim$150~K on cooling and are continuously suppressed
below $T\sim$150~K, showing no anomaly across $T_N$. 
Currently, we do not understand why there
is no anomaly in the spin correlation lengths at 150 K 
[see Figs. 4(a) and 4(b) Insets].

\section{Discussion and conclusion}

In general, CO/OO ordering is strongly coupled to the lattice, 
has a large energy scale, and occurs at 
higher temperatures than magnetic ordering. As a consequence,
the development of magnetic order at low temperature usually has no effect on the lattice distortions induced
by the CO/OO 
order. For previously studied half-doped
manganites \cite{Radaelli,jirak,aladine,ye,sternlieb,Larochelle2},
orbital ordering is always established simultaneously with charge 
ordering \cite{murakami,Dhesi}. In addition, the CE-type AF order occurring at 
low temperatures
stabilizes the CO/OO ordered phase and the orthorhombicity of the system
saturates below $T_N$ \cite{Radaelli}.  
Since PCMO has a smaller $A_{0.5}A^{\prime}_{1.5}$ ionic radius and 
larger lattice distortion than that of LSMO, one would
expect CO/OO order in PCMO to be more robust than the magnetic order.
Instead, the dramatic reduction of the orthorhombicity near $T_N$ indicates 
a strong spin-lattice coupling that can influence the distortion already 
established by 
CO/OO ordering. At present, it is unclear why PCMO should 
behave differently from other half-doped manganites. 
Perhaps the
small Pr/Ca ionic sizes and large lattice distortion in this
material can enhance the CE-type AF superexchange
interaction and make it comparable to the energy of
CO/OO ordering. For LSMO,  
inelastic neutron scattering experiments \cite{senff} have shown that 
 the ferromagnetic exchange coupling along the
zigzag chain direction [see Fig. 1(a)] is about 5.5 times larger than that of 
AF exchange in between the chains
($J_{FM}/J_{AF}=9.98\ {\rm meV}/1.83\ {\rm meV}\approx 5.5$).

In the case of PCMO, our preliminary spin wave measurements based on the Hamiltonian
similar to that reported in ref. \cite{senff} suggest 
that this ratio becomes $J_{FM}/J_{AF}=8.7\ {\rm meV}/6.5\ {\rm meV}\approx 1.34$ \cite{chi}.  
Therefore,  the AF exchange interaction is much stronger in PCMO than in LSMO, 
making a more robust AF CE structure with little anisotropy between $a$ and $b$ axis directions.
This means that the AF order in PCMO prefers a tetragonal structure
rather than orthorhombic \cite{Goodenough}, and provides a competing energy scale to the 
already established CO/OO ordering. 
In any case, our data clearly indicate that
the magnetic exchange energy in PCMO is an important
competing force and must be taken into account to understand its low temperature 
electronic properties. Furthermore, the spin-lattice coupling in PCMO is much stronger than that for
other half-doped manganites.

In summary, we have carried out neutron scattering studies of the lattice and magnetic structure
of the layered half-doped manganite PCMO. 
The system first displays CO/OO order and then develops 
CE-type AF order at low temperatures.  We have discovered that AF order can have a large effect on the
already established lattice distortions induced 
by the CO/OO.  This result indicates a strong spin-lattice coupling in PCMO. 
It also contrasts with all other known half-doped manganites, where AF order has little or
no influence on orthohombic strains in the system.  We argue that the reason for this difference is
because magnetic exchange coupling in PCMO is much more isotropic, favoring a tetragonal AF crystal structure.
As a consequence, the low-temperature electronic properties of the half-doped manganites are 
determined not only by CO/OO ordering, but are also affect by strong spin-lattice coupling.

\section {Materials and Methods}
We grew single crystals of PCMO using the traveling solvent floating
zone technique. High resolution neutron powder diffraction (HRNPD)
experiments were carried out on BT-1 at the NIST Center for Neutron
Research (NCNR) with powder of crushed single crystals. Elastic
neutron scattering measurements were carried out on the thermal
triple-axis instruments BT-7 and BT-9 at NCNR.  
Rietveld analysis on
the powder data indicates that the crystals were single phase without
detectable impurities.
The crystals were mounted in a
closed cycle He displex and aligned in successive
orientations to allow the wavevectors in the form of $(h, h, l)$,
$(h, k, 0)$ and $(h, 0, l)$ accessible in the horizontal scattering
plane.  Neutron energies of 14.7 meV and 13.7 meV were used with 
pyrolytic graphite crystals as monochromator, analyzer and
filters. 

\begin{acknowledgments}
We thank D. Khomskii for helpful discussions.
The work was supported by U.S. NSF-DMR0453804.
ORNL is managed by UT-Battelle, LLC, for the U.S. Dept. of Energy
under contract DE-AC05-00OR22725. This work was also performed under
the US-Japan Cooperative Program on Neutron Scattering
\end{acknowledgments}




\begin{thebibliography}{}
\bibitem{distortion} The average $A$-site ionic radius for LSMO is 
$r_{ave}=1.2865$ \AA, while it is $r_{ave}=1.1798$ \AA\ for PCMO. This is a rather large 
reduction.
\end{thebibliography}
\end{document}